\documentclass[aps]{revtex4}

\usepackage{epsfig,graphicx,subfigure,amsmath}

\begin{document}
\newcommand{\be}{\begin{equation}}
\newcommand{\ee}{\end{equation}}
\newcommand{\bea}{\begin{eqnarray}}
\newcommand{\eea}{\end{eqnarray}}
\newcommand{\nn}{\nonumber}
\newcommand{\ba}{\bea \begin{array}}
\newcommand{\ea}{\end{array} \eea}
\renewcommand{\(}{\left(}
\renewcommand{\)}{\right)}
\renewcommand{\[}{\left[}
\renewcommand{\]}{\right]}
\newcommand{\bc}{\begin{center}}
\newcommand{\ec}{\end{center}}
\newcommand{\p}{\partial}

\newcommand{\red}{\textcolor{red}}
\newcommand{\mb}[1]{ \mbox{\boldmath$#1$}}
\newcommand{\ds}{\displaystyle}
\newcommand{\beq}{\begin{eqnarray}}
\newcommand{\eeq}{\end{eqnarray}}
\newcommand{\beqq}{\begin{eqnarray*}}
\newcommand{\eeqq}{\end{eqnarray*}}

\newcommand{\g}{\gamma}
\newcommand{\epsv}{\epsilon}
\newcommand{\eps}{\varepsilon}

\newcommand{\x}{\mbox{\boldmath$x$}}
\newcommand{\hx}{\mbox{$\hat x$}}
\newcommand{\n}{\mbox{\boldmath$n$}}
\newcommand{\J}{\mbox{\boldmath$J$}}
\newcommand{\y}{\mbox{\boldmath$y$}}
\newcommand{\z}{\mbox{\boldmath$z$}}

\title{The Gated Narrow Escape Time for molecular signaling}
\author{J\"urgen Reingruber and David Holcman}
\affiliation{Department of Computational Biology, Ecole Normale Sup\'erieure, 46 rue
d'Ulm 75005 Paris, France.}
\begin{abstract}
The mean time for a diffusing ligand to activate a target protein located on the surface of a
microdomain can regulate cellular signaling. When the ligand switches between various states
induced by chemical interactions or conformational changes, while target activation occurs in only
one state, this activation time is affected. We investigate this dynamics using new equations for
the sojourn times spent in each state. For two states, we obtain exact solutions in
dimension one, and asymptotic ones confirmed by Brownian simulations in dimension 3. We find that
the activation time is quite sensitive to changes of the switching rates, which can be used to
modulate signaling. Interestingly, our analysis reveals that activation can be fast although the
ligand spends most of the time 'hidden' in the non-activating state. Finally, we obtain a new
formula for the narrow escape time in the presence of switching.
\end{abstract}

\maketitle
Cellular chemical reactions depend on the activation of small targets by diffusing ligands. This
process can be regulated by several parameters such as the geometry of the cellular microdomain,
the target shape, or the state of the target and ligand upon encounter. {In the past, the
activation rate and survival probability of the target were studied for diffusion-influenced
chemical reactions where the reactivities of the target or the ligands stochastically fluctuate in
time \cite{ZhouSzabo_JPC1996,Berezhkovskiietal_PRE1996,Doering_LectureNotes_2000}. Interestingly, for slow gating dynamics it was found that both processes are not equivalent, and, for example, a gated
target may lead to non exponential behaviour at short times
\cite{Zwanzig_JCP1992,SzaboSchultenSchulten1980,ZhouSzabo_JPC1996,Szaboetal_1982}.} When the target
has a fluctuating potential barrier, there is an optimal combination of parameters for which the
mean activation time is minimal, leading to a resonant-like phenomena
\cite{DoeringGadooua_PRL1992,BierAstumian_PRL1993}. In cellular microdomains, target activation
may depend on the state of the ligand. This is for example the case in the cytoplasm, where enzymes
can switch between an inactive and active state, or in the nucleus, where a transcription factor
needs to be first activated in order to bind to a specific DNA promoter
\cite{BookPtashne,BensaudeRev_BioCell2008}. Recently, intermittent search scenarios were introduced
and extensively analyzed for a dynamics switching stochastically between fast ballistic phases and
slow diffusive phases, while the target can only be found in the diffusing phase
\cite{Benichou_Switch_PRL2005,Benichou_Switch_JPCondM2005,Benichou_Switch_NatPhys2008}:
interestingly, switching can decrease the search time, and there are optimal search strategies that
minimizes the search time.

In the absence of switching, target activation is determined by the narrow escape time (NET), which
is the mean time for a Brownian ligand to find a small target in a confined environment
\cite{Ward1,Ward4,Grigorievetal2002,HolcmanetalNE3,HolcmanPNAS2007}. The NET has been used to
compute the mean and variance of chemical reactions with few molecules diffusing in microdomains
\cite{HolcmanSchuss2005_JCP}, to estimate the probability and the arrival time of viral particles
to nuclear pores \cite{LagacheDautyHolcman_PRE2008}, or to study the early steps of
phototransduction in rod photoreceptor \cite{ReingruberHolcman_PRE2009}.

We study in this letter the gated narrow escape time (GNET) to exit the domain $\Omega$ through a
small window, if a diffusing ligand stochastically switches between two states 1 and 2 with
diffusion coefficients $D_1$ and $D_2$, and can exit only in state 1. Switching may be due to
conformational changes or chemical interactions. We estimate the GNET using new equations for the
sojourn times the ligand spends in the different states. We find that switching not only affects
drastically the exit time and the sojourn times, but also, only for $D_2>D_1$ the GNET can be
optimized as a function of the switching rates. In addition, a ligand may exit almost as fast as
possible, although it spends most of the time in state 2. Finally, we give a new formula for the
GNET in dimension three, which extends the NET formula to the switching case. We also discuss
briefly possible applications in cellular signaling.
\begin{figure}[h!]
\begin{center}
      \includegraphics[scale=0.5]{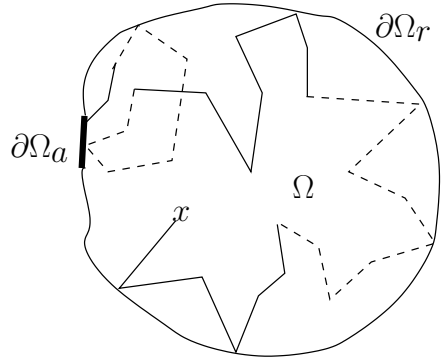}\hspace{0.cm}
       \caption{Example of a trajectory of a diffusing Brownian ligand in a confined
       domain $\Omega$ and randomly switching between two states 1 (continuous line)
       and 2 (dashed line). In state 2, the ligand is reflected all over the boundary,
       while in state 1 it is absorbed at $\p \Omega_a$.}
  \label{fig_switch}
\end{center}
\end{figure}

\begin{figure}[h!]
\begin{center}
      \includegraphics[scale=0.8] {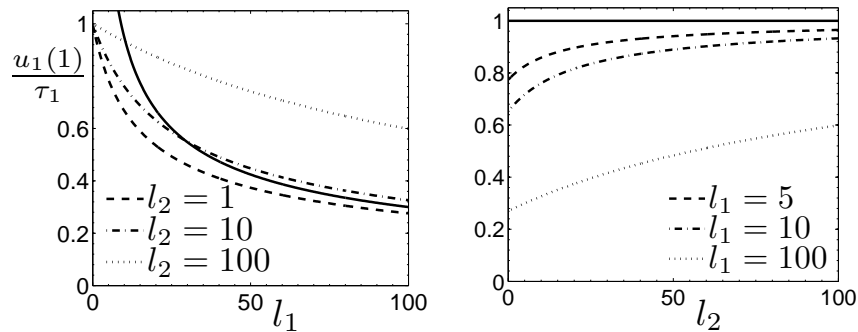}\hspace{0.cm}
       \caption{Graph of the sojourn time $u_1(1)$ obtained from eq.~\ref{appva2}
       as a function of $l_1$ and $l_2$, scaled by the mean first passage time $\tau_1$.
       In the left panel, the continuous curve is the asymptotic $3/\sqrt{l_1}$
       for $l_1\gg1$ and $\sqrt{l_1}\gg l_2$.}
\label{fig_v1}
\end{center}
\end{figure}

\begin{figure}[h!]
\begin{center}
      \includegraphics[scale=0.8]{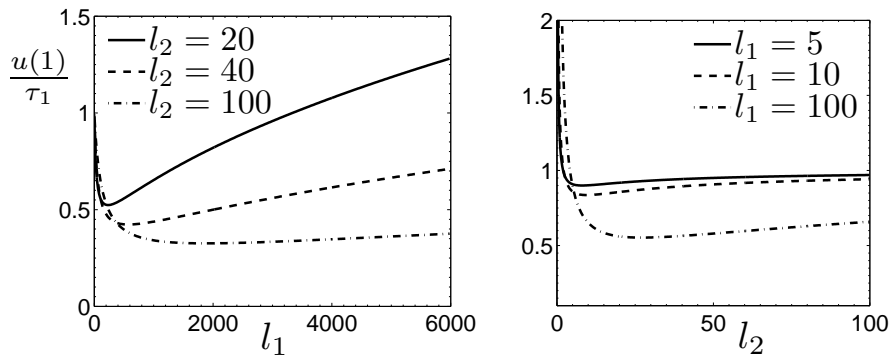}\hspace{0.cm}
       \caption{Graph of the exit time $u(1)$ from eq.~\ref{eqhatu1}
      as a function of $l_1$ and $l_2$ for $\kappa=0.1$, scaled by $\tau_1$.}
  \label{fig_u1}
  \vspace{-0.5cm}
\end{center}
\end{figure}

\begin{figure}[h!]
\begin{center}
      \includegraphics[scale=0.8]{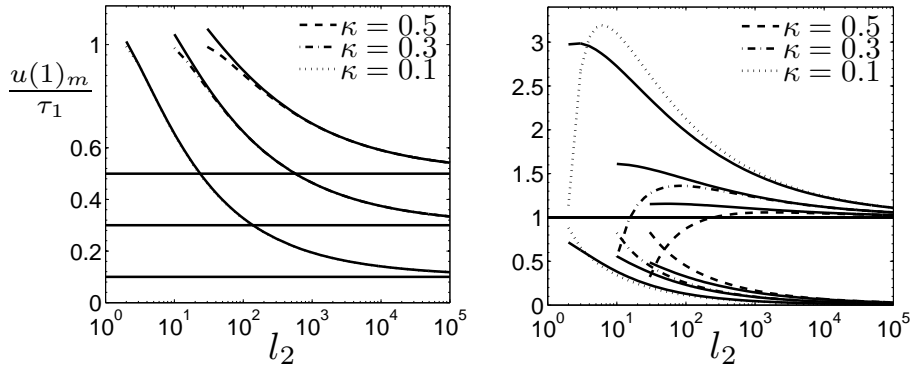}\hspace{0cm}
       \caption{The left panel displays the minimum $u(1)_{m}$ scaled by $\tau_1$
       as a function of $l_2$ and $\kappa$. The horizontal lines are ${\kappa}$, and the continuous curves are calculated
       using eq.~\ref{asympmin}. The graphs do not extend up to $l_2=0$ because a minimum exists only for
       $l_2>\tilde l_2(\kappa)$ (see text). The right panel shows the ratio $u_1(1)_m/\tau_1$
        (curves below the horizontal line) and $u_2(1)_m/\tau_2$ (curves above)
        of the corresponding sojourn times. The continuous curves are calculated using eq.~\ref{asympmin}. }
      \vspace{-0.5cm}
  \label{fig_minu1}
\end{center}
\end{figure}

A Brownian ligand diffuses in a confined domain $\Omega$ while switching between states 1 and 2
with Poissonian rate constants $k_{12}$ and $k_{21}$ and diffusion constants $D_1$ and $D_2$. Upon
encounter, the ligand activates the target located on the small boundary portion $\p \Omega_a$ only
in state 1, while it is reflected otherwise everywhere on the boundary. To study the GNET, we
consider the mean sojourn times $u_n(\x,m)$ the ligand spends in state $n$ before exiting,
conditioned on starting at position $\x$ in state $m$. From the backward Chapman-Kolmogorov
equation \cite{BookSchuss,BookRisken} we find that the $u_n(\x,m)$ satisfy the coupled system of
equations
\beq\label{coupledEq}
L^*_m u_n(\x,m) - \sum_{i=1}^2 k_{mi} (u_n(\x,m) -u_n(\x,i)) = -\delta_{nm},
\eea
where $L^*_m$ is the backward Kolmogorov operator in state $m$, which in our case is $L^*_m=D_m
\Delta$, and we have absorbing boundary conditions on $\p \Omega_a$ for $u_n(\x,1)$, and reflecting
conditions otherwise. Eq.~\ref{coupledEq} separately constitutes a closed system of equations for
each value $n$, and it is sufficient to study the equations for $n=1$, because the solutions
$u_2(\x,m)$ are obtained through the linear transformation
\beq\label{transformabsrefl}
\begin{pmatrix} u_2(\x,1) \\ u_2(\x,2) \end{pmatrix} = \frac{k_{12}}{k_{21}}
\begin{pmatrix}  1 & 0 \\ 0 & 1 \end{pmatrix}
\begin{pmatrix} u_1(\x,1) \\ u_1(\x,2) \end{pmatrix}
+ \begin{pmatrix} 0 \\ k_{21}^{-1}\end{pmatrix}\,.
\eeq
The mean sojourn times $u_1(1)$, $u_1(2)$, $u_2(1)$ and $u_2(2)$ are obtained by averaging
eq.~\ref{coupledEq} and eq.~\ref{transformabsrefl} over an initially uniform distribution,
satisfying $ u_1(2) = u_1(1)$, $u_2(1)=u_1(1) {k_{12}}/{k_{21}}$ and $u_2(2) = u_2(1) +
{1}/{k_{21}}$. The mean times $u(1)$, $u(2)$ and $u$ to exit the domain starting uniformly
distributed in state 1, 2, and in state 1 and 2 with equilibrium probability
$(p_1,p_2)=(\frac{k_{21}}{k_{12}+k_{21}},\frac{k_{12}}{k_{12}+k_{21}})$, are
\bea
u(1) &=& u_1(1) + u_2(1) = u_1(1) \( 1 +{k_{12}}/{k_{21}}\) \label{uArefecting}\\
u(2) &=& u_1(2) + u_2(2) = u(1) + {1}/{k_{21}} \label{uBrefecting} \\
u    &=& p_1 u(1) + p_2 u(2) = u(1) + \frac{k_{12}}{k_{21}(k_{12} + k_{21})} \,.\label{urefecting}
\eea
{\it \noindent The GNET in one dimension}\\
We now solve the GNET problem in one dimension where $\Omega$ reduces to the interval $0 \le x \le
L$ with an absorbing boundary at $x=0$ in state 1, and a reflecting boundary at $x=L$. The
one-dimensional analysis already reveals many features that remain valid also in higher dimensions
where only asymptotic results are available. Using the scaled parameters
\bea
l_1={k_{12} L^2}/{D_1}\,, \quad l_2={k_{21} L^2}/{D_2}\,, \quad \kappa={D_1}/{D_2}
\eea
we solve eq.~\ref{coupledEq} and eventually obtain
\bea
u_1(x,1)&=& \frac{L^2}{D_1}\frac{ l_1\(
\cosh \sqrt{l_1+l_2} -  \cosh (\sqrt{l_1+l_2} \frac{L-x}{L})\)}{ (l_1+l_2)\sqrt{l_1+l_2}\sinh \sqrt{l_1+l_2}}\nn\\
&& + \frac{l_2}{l_1+l_2} \frac{(2L- x)x}{2 D_1}\,,
\eea
from which we derive for the averaged sojourn time
\bea
u_1(1) = \tau_1 -  \frac{l_1}{l_1+l_2} \( \tau_1 - \frac{L^2}{D_1} f(l_1 +l_2\)
\,,\label{appva2}
\eea
where $\tau_1= {L^2}/{(3D_1)}$ is the mean first passage time to exit in state 1 without switching,
and $f(x) = \frac{\coth\sqrt{x}}{\sqrt{x}} -\frac{1}{x}$ is monotonically decreasing from ${1}/{3}$
at $x=0$ towards 0 for $x\to \infty$. In fig.~\ref{fig_v1}, we plot $u_1(1)$ as a function of $l_1$
and $l_2$. Interestingly, because $u_1(1) \le \tau_1$, we found the non-intuitive result that the
sojourn time the ligand spends in state 1 before exiting is always smaller than the mean first
passage time $\tau_1$ to exit in this state without switching. Even more surprising, $u_1(1)$ can
become arbitrarily small by increasing $l_1$ (resp. the switching rate $k_{12}$). This behavior can
be understood as follows ({see also \cite{Doering_LectureNotes_2000}}): for a ligand that starts
uniformly distributed in state 1, the probability to find it in the neighborhood of the absorbing
boundary at $x=0$ decreases quickly. But, in state 2 the probability distribution is re-homogenized
and after switching back to state 1, the density around $x=0$ is higher compared to the
non-switching case leading to a faster exit.
Because the sojourn time $u_1(1)$ can be arbitrarily small, we
wonder how this affects the mean time $u(1)$ to exit when starting
initially in state 1. Using eq.~\ref{appva2} we obtain
\bea\label{eqhatu1}
u(1) = \tau_1  + \frac{l_1}{l_2}\( \frac{L^2}{D_1} f(l_1+l_2) +(\kappa-1) u_1(1) \),
\eea
which shows that $u(1)\ge \tau_1$ for $\kappa \ge 1$. Obviously, by
switching to a state with a smaller diffusion constant, we cannot
speed up exit. However, when $\kappa <1$, interestingly, the
situation changes: for fixed $l_2$, by expanding $u(1)$ in
eq.~\ref{eqhatu1} as a function of $l_1$ for $l_1\ll 1$,
we find that $u(1)$ initially decreases if $l_2>\tilde l_2(\kappa)$, where $\tilde l_2(\kappa)$ is
the root of $f(l_2) +(\kappa-1)/3=0$. Together with the asymptotic $u(1)\sim \sqrt{l_1}$ for large
$l_1$, we conclude that $u(1)$ has a minimum $u(1)_{m}<\tau_1$ as a function of $l_1$ for
$l_2>\tilde l_2(\kappa)$. Similar to the results of \cite{Benichou_Switch_PRL2005}, we found here
that the exit time $u(1)$ is minimal for a certain value of $k_{12}$ (that depends on $l_2$ and
$\kappa$). The left panel of fig~\ref{fig_u1} shows $u(1)$ as a function of $l_1$ for various
$l_2$, and the right panel displays $u(1)$ as a function of $l_2$ for various $l_1$, showing that
$u(1)$ also has a minimum as a function of $l_2$.
For given values $l_2$ and $\kappa<1$, we now study the minimum $u(1)_{m}$ of $u(1)$ achieved at
$l_1=l_{1,m}$. The left panel of fig.~\ref{fig_minu1} shows $u_{1,m}$ scaled by $\tau_1$ as a
function of $l_2$ for various $\kappa$ ($l_{1,m}$ is found by numerically solving $\p u(1)/\p
l_1=0$). An asymptotic analysis for large $l_2$ reveals that the position $(l_{1,m},l_2)$ of the
minimum is defined by
\bea\label{asympmin}
\frac{l_2^2}{l_{1,m}^{3/2}}= \frac{3\kappa}{2 (1-\kappa)}\,,
\eea
from which we derive for large $l_2$ the asymptotic behaviour $u_1(1)_m = O(l_2^{-1/3})$,
$u_2(1)_m=\tau_2 +O(l_2^{-1/3})$ and $u(1)_m=\tau_2 + O(l_2^{-1/3})$, where $\tau_2=\kappa \tau_1$
is the mean first passage time for a ligand diffusing with a diffusion constant $D_2$. The right
panel of fig.~\ref{fig_minu1} displays the ratio $u_1(1)_m/\tau_1$ and $u_2(2)_m/\tau_2$ as a
function of $l_2$ for different $\kappa$. Surprisingly, when exit is fast and $u(1)$ approaches its
lower limit $\tau_2$, the ligand spends almost all its lifetime in state 2 where it cannot exit.
When considering $u(1)$ as a function of $l_2$ for a given value of $l_1$, the minimum position
$(l_1,l_{2,m})$ differs from $(l_{1,m},l_2)$, however, it satisfies a very similar relation
compared to eq.~\ref{asympmin}. In general, for $\kappa <1$, $u(1)$ has no local minimum as a
function of $(l_1,l_2)$, but is lower bounded by $\tau_2$. Only for $\kappa\ge 1$ we have the
global minimum $u(1)_m=\tau_1$ trivially attained for $l_1=0$ when the ligand does not switch and
stays in state 1. Furthermore, for $\kappa=1$ there is a strategy to exit in almost minimal time
$\tau_1$ while spending most of the time hidden in the state 2 where exit is not possible. With
$\alpha \ll 1$, the strategy is to choose $\alpha \sqrt{l_1} \gg 1$ and $l_2=\alpha
l_1$ such that $u(1)\approx \tau_1$, and since $u_1(1)/u_2(1)= \alpha$, it follows that the ligand
spends only the small fraction $\alpha/(1+\alpha)$ of its time in state 1.

In summary, the lower limit of $u(1)$ corresponds to a ligand diffusing all the time with the
maximal diffusion constant, and interestingly, fast exit can be achieved even when diffusing most
of the time in the state where exit is not possible. Furthermore, for $\kappa<1$, $u(1)$ has a
minimum as a function of $l_1$ resp. $l_2$ for non-vanishing switching rates, however, as shown in
figure \ref{fig_u1}, the graph for $u(1)$ around and past the minimum is quite flat, while it
decays steeply for small switching rates. Thus, the behaviour at small rates is an efficient
mechanism to modulate the activation time, and thus cellular signaling.

\begin{figure}[h!]
\begin{center}
      \includegraphics[scale=0.8]{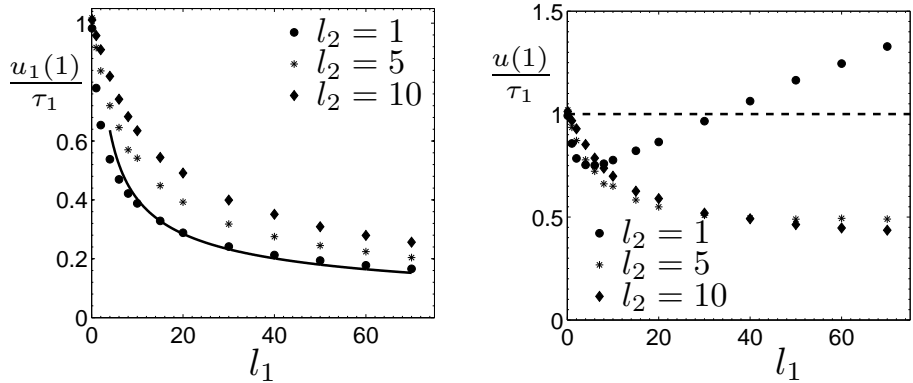}\hspace{0.cm}
 \caption{Simulation results for the sojourn time $u_1(1)$ and the exit time $u(1)$
 as a function of $l_1$ for different $l_2$ (marked by various symbols). The results are scaled by
 the NET $\tau_1$.
 The continuous line in the left panel is the asymptotic $\frac{4}{\pi \sqrt{l_1}}$ obtained from eq.~\ref{3du11}
 (see also text). The results for $u(1)$ in the right panel are obtained for $\kappa=0.1$.}
  \label{fig_3d}
\end{center}
\end{figure}

{\it \noindent The GNET in a 3-dimensional microdomain} \\
We now extend our previous analysis to a ligand confined in a 3-dimensional domain $\Omega$ that
activates upon hitting a small boundary patch $\p\Omega_a$ in state 1. Without switching, the time
reduces to the narrow escape time (NET)
\cite{Ward4,HolcmanetalNE3,Grigorievetal2002}. The NET analysis
showed that outside a small boundary layer around the absorbing hole
of radius a few $a$, where $a$ characterizes the extent of
$\p\Omega_a$, the positional NET is almost independent of the
initial ligand position
\cite{HolcmanetalNE3}.
Using the scaling introduced in \cite{ReingruberAbadHolcman}, we define the dimensionless variables
and functions $\hat
\x={\x}/{a}$, $l_1={k_{12}a^2}/{D_1}$, $l_2= {k_{21}a^2}/{D_2}$, $v_1(\hat
\x) =  \frac{a D_1}{|\Omega|}u_1(\x,1)$, $v_2(\hat \x) = \frac{a D_1}{|\Omega|}u_1(\x,2)$,
and obtain from eq.~\ref{coupledEq} the scaled equations
\beq \label{scaledeq3dim}
\begin{array}{l}
\Delta v_1(\hat \x) - l_1 (v_1(\hat \x)- v_2(\hat \x)) = -|\hat \Omega|^{-1} \\
\Delta v_2(\hat \x) + l_2 (v_1(\hat \x)- v_2(\hat \x)) = 0\,,
\end{array}
\eeq
where $|\hat \Omega|=|\Omega|/a^3 \gg 1$. The boundary conditions
are absorbing on $\p
\hat \Omega_a$ for $v_1(\hat\x)$, and otherwise reflecting. The solutions of
eq.~\ref{scaledeq3dim} for general values $l_1$ and $l_2$ are not at hand, thus we present here
asymptotical results that clarify the effect of switching. For $l_1 \ll 1$ or $l_2\gg l_1$,  at
leading orders, $v_1(\hat \x)$ is solution of the NET problem $\Delta v_1(\hat \x)  = -|\hat
\Omega|^{-1}$. When $l_2 \ll \sqrt{l_1}$, the leading order solution for $v_1(\hat\x)$ is found by solving $\Delta
v_1(\hat \x) - l_1 (v_1(\hat \x)-v_1) = -|\hat \Omega|^{-1}$, where
$v_1$ is the spatial average of $v_1(\hat \x)$. Using the asymptotic
solutions for $v_1(\hat \x)$, the leading order expression of the
sojourn time $u(1)= (1+k_{12}/k_{21})\frac{|\Omega|}{a D_1}v_1$ is
\bea \label{3du11}
u(1) = (1+\frac{k_{12}}{k_{21}})\left\{
\begin{array}{l} \displaystyle
\tau_1  \,,  \quad \,l_1 \ll 1  \mbox{  or  } l_2 \gg l_1   \\
 \displaystyle  \frac{|\Omega|}{|\p \Omega_a| \sqrt{D_1 k_{12}}}, \,
l_1 \gg 1\,, \sqrt{l_1} \gg l_2\,.
\end{array}\right.
\eea
To confirm the behavior of $u(1)$ and $u_1(1)$ as a function of $l_1$ and $l_2$, we used Brownian
simulations together with the Gillespie-algorithm \cite{Gillespie_JCompPhys1977} to model switching
in a sphere of radius $r=30$ with a circular hole of radius $a=1$. In the left panel of
fig.~\ref{fig_3d}, we show simulation results for $u_1(1)$ as a function of $l_1$ for various
$l_2$. We obtain that $u_1(1)\le \tau_1$, and confirm the asymptotical behaviour $u_1(1)/\tau_1
\approx 4/(\pi\sqrt{l_1})$ which follows from eq.~\ref{3du11} by using $\tau_1 \approx
|\Omega|/(4aD_1)$ \cite{HolcmanPNAS2007} and $|\p \Omega_a|=\pi a^2$. In the left panel of
fig.~\ref{fig_3d} we display the simulations results for $u(1)$ as a function of $l_1$ for various
$l_2$ and $\kappa=0.1$ ($D_1=1,D_2=10$). The plot shows that $u(1)$ has a minimum smaller than
$\tau_1$ which is attained at some value $l_{1,min}>0$. For $\kappa \ge 1$, we have $u(1)\ge\tau_1$
(not shown).

The asymptotic expressions in eq.~\ref{3du11} correspond to two different physical regimes (see
also the regimes discussed in \cite{Doering_LectureNotes_2000}): In the range where $u_1(1)\approx
\tau_1$, the GNET is the NET $\tau_1$ divided by the probability
$p_1=\frac{k_{21}}{k_{12}+k_{21}}$ to find the ligand in state 1, which indicates a mean-field
situation where switching and absorption proceed independently. In this range the switching
dynamics can be approximated by an effective non-switching diffusion process with diffusion
constant $D_{eff}=D_1/p_1$. In the case of $\sqrt{l_1}\gg l_2, l_1 \gg 1$, $u(1)$ and $u_1(1)$ are
inversely proportional to the surface of the target, similar to the reaction-controlled NET to a
partially absorbing hole \cite{ReingruberAbadHolcman,Doering_LectureNotes_2000}. This is very
different from the mean-field situation and indicates the appearance of strong correlations.

To conclude this study of the mean time to activate a target by a switching Brownian ligand, we
found the new formula eq.~\ref{3du11} extending the NET to the case of switching. For $D_2 >D_1$,
switching can significantly decrease the exit time compared to non-switching. For small values of
$l_1$ and $l_2$, the mean exit time is very sensitive to changes in these parameters (see
fig.~\ref{fig_3d} and fig.~\ref{fig_u1}), and this property can be exploited to modulate cellular
signaling. For example, if switching is due to a chemical reaction, the value of $l_1$ is changed
via $k_{12}$ by modulating the concentration of the reaction partner.

We finish by giving two examples in signal transduction where switching might be important.
The first is related to the search time for a promoter DNA-site by a transcription factor (TF),
which alternates between a 3-dimensional diffusion in the nucleus and one-dimensional diffusion
along the DNA \cite{Elf_Science2007,AustinCox_PRL2006,HippelBerg_JBC1989}. We consider a TF
diffusing along the DNA and switching between two states, triggered by stochastic conformational
changes of the TF. In state 1, diffusion is slow due to a high affinity for the DNA where the TF carefully scans the DNA base pairs. In state 2, diffusion is much faster, but
the TF does not accurately scan the DNA. The search time of this can be much reduced
compared to a TF which carefully checks all the DNA base pairs. The second application illustrates
our finding that a diffusing ligand can activate a target fast, although it may spend almost all
its time in a state where it has no affinity for the target. This is relevant if a ligand needs to
activate a target in a state where it is also prone to some degradation. We found that a ligand can
largely avoid degradation and still perform fast target activation by switching between two states,
such that it spends most of its time in state 2, where it cannot be degraded nor activate the
target.

\bibliographystyle{apsrev}

\end{document}